\documentclass{article}

\usepackage{arxiv}

\usepackage{graphicx}
\usepackage{cite}
\usepackage{graphicx}
\usepackage{amssymb}
\usepackage{mathrsfs}
\usepackage[toc,page]{appendix}
\usepackage{algorithm}
\usepackage{algorithmic}
\usepackage{multicol}
\usepackage{hyperref}
\usepackage{color}

\makeatletter

\newcommand{\beit}{\begin{itemize}}
	\newcommand{\eit}{\end{itemize}}

\newsavebox{\savepar}

\title{\huge PACLP: A fine-grained Partition-based Access Control Policy Language for Provenance}

\author{Xinyu Fan\thanks{Corresponding author}, Faen Zhang,  Jianfei Song, Jingming Guo, Fujie Gao\\
fanxinyu@ainnovation.com \{zhangfaenainnovation, songjianfeiainnovation\}, \\
\{guojingmingainnovation, gaofujieainnovation\}@gmail.com\\
AInnovation Technology Ltd.} 

\begin{document}
\maketitle

\begin{abstract}
	Even though the idea of partitioning provenance graphs for access control was previously proposed, employing segments of the provenance DAG for fine-grained access control to provenance data has not been thoroughly explored. Hence, we take segments of a provenance graph, based on the extended OPM, and defined use a variant of regular expressions(based on Xpath query language) to determine which set of nodes can be accessed,and utilize these methods in our fine-gained access control language. It can not only return partial graphs to answer access requests but also introduce segments as restrictions in order to screen targeted data. 
	
\end{abstract}
\section{Introduction}

Data provenance logs historical operations performed on documents. Provenance can be expressed as a directed acyclic graph (DAG),  illustrating how a data artifact is processed by an execution. In such a DAG of provenance under the Open Provenance Model (OPM)\cite{MoreauCFFGGKMMMPSSB11}, nodes present three main entities including \emph{Artifact}, \emph{Agent} and \emph{Process} and edges represent connections to the main entities. Provenance Access Control is considered a crucial research topic for big data security. Because users can request and be granted access to files and sources separately,the sensitivity of files and their provenance can be different. In some situations, provenance itself may consist of sensitive information which might require more protection than its attached document. For instance, although a programming project can be published to the public, its authors and executed operations should be kept as a secret, to prevent leaking the techniques. Therefore, access control  to the provenance data itself is required, so that it not only allows eligible users to access the provenance data, but also protects it from unauthorized access. Privacy and security of provenance are perceived as the main bottleneck to broad applications of provenance \cite{McDaniel11}\cite{BhuyanLRAZ18}\cite{Gonzalez-Manzano17}. 

Privacy helps individuals maintain their autonomy and individuality, and security is the protection from theft and damage to provenance, as well
as from disruption or misdirection of provenance. Hence, lacking protection for privacy and security can not convince users to trust the application of provenance. 

Although there has been great progress in provenance access control, there still exist some difficulties. Existing traditional access control techniques including role-based access control\cite{SandhuS94}, attribute-based access control \cite{KerrF16} cannot be applied to provenance access control straightforwardly, because of provenance is a type of meta-data with a specific data structure. Due to the diversity and particularity of provenance data structure, traditional access control technology is not suitable for provenance access control. Therefore, there is an urgent need to develop appropriate source access control languages. How to define fine-grained access control policies under a proper provenance model is the main concern of this paper.

Previous provenance DAG access control language has not employed different types of segments in provenance graphs thoroughly. Although Danger et al. came up with the idea of returning origin subgraphs by answering access queries, their approach relied on explicitly enumerated node sets and could not cope with scenarios with large Numbers of nodes and unpredictable nodes. Therefore, a mechanism is proposed in this chapter to define a collection of nodes by a summarizing approach. For example, a string of connected nodes can be defined in a policy by nominating a starting node and an ending node. More concretely, a provenance logs an assignment written by a student and the operation the student carries out upon the assignment. The provenance also records that the student submitted the assignment to a professor and actions from the professor to grade it, attach comments, revisions, \emph{etc.} However, students are not allowed to gain any information about the operations carried out by the markers. However, it cannot be known in advance, for a given submissions, how many operations will be involved in the marking. A professor may access the submission multiple times and it may even be marked by more than one staff members. Therefore, a policy is required that can block access to a portion of the provenance from the point of submission until the assignment is returned to the student by the professor, regardless of how many processes have been done between those points.

In this paper, we propose a Segment-based Access Control Policy Language for Provenance (PACLP) by extending existing policy languages. PACLP enables the specification of partial provenance graphs as well as transformation scope, transformation mode, and transformation labels, in order to partition a provenance graphs in a fine-grained approach and define how to transform the collected nodes into a new graph.  And our major contributions can be summarized as follows.
\begin{itemize}
	\item The attributes are stored to each node in a provenance graph to support more fine-grained access control policy, which can also be used in policies to specify which nodes are the targets of the policy. 
	\item We use regular expressions , which based on Xpath query language, to define multiple ways to partition provenance DAG, including single node, node type, path, and subgraph, which could be utilized as conditions for access control policy.
	
\end{itemize}
\section{Related Works}
Several previous works have made significant contributions to provenance access control. Ni \emph{et.al} \cite{NiXBSH09} proposed an initial fine-grained provenance access control language based on XACML syntax, which was customized according to the provenance model of recording operational attributes.  Subsequently, several papers proposed access control languages of the  DAG provenance models, and tried to partition provenance graphs and return partial graphs to users\cite{ChenENN15}\cite{DangerCMB15}. In addition, SPARQL query templates were presented to answer queries that record attributes in their provenance. 

Cadenhead \emph{et al.}\cite{CadenheadKKT11} extended an existing provenance access control language proposed by Ni \emph{et al.}\cite{NiXBSH09}, which introduced regular expression to protect traditional data items as well as their relationships from unauthorized users. It was an XML-based structure policy language and associates grammar based on provenance graphs. In order to evaluate the effectiveness of their policy language, a prototype language based on their architecture utilizing Semantic Web technologies has been implemented. 

Danger \emph{et al.}\cite{DangerCMB15} proposed a method that allows policies to define subgraphs that can be transformed through three levels of abstractions, which presented an algorithm for transforming provenance graphs and generating accessible versions of queries.  Although Danger \emph{et.al} presented the idea of returning provenance sub-graphs by answering access queries, their approach relies upon explicitly enumerated node-sets. This approach is not effective when the policies contains a large amount of nodes. In addition, in some scenarios, the set nodes in a collection can not be predicted in advance. \\

\section{Provenance Access Control}

\subsection{The Basics of the policy language}

\begin{figure}[thb]
	\vspace{-0.5cm}
	\centering
	\includegraphics[scale=0.35]{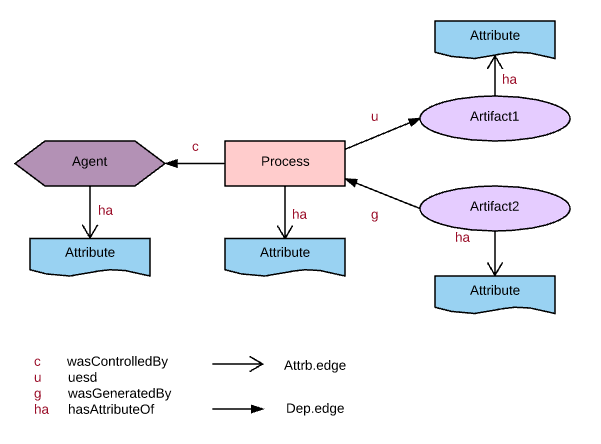}
	\vspace{-0.5cm}
	\caption{The extended version of OPM: OPM$^+$ Schema} 
	\label{fig2}
\end{figure} 

Given the OPM$^+$ $<$T, L, G$>$, an OPM$^+$ instance is defined by a provenance graph $G_i$ = $<V_i, E_i>$, where $V_i$ is a set of entities and $E_i$ $\in V_i \times V_i \times L$.  $\tau$: $V_i \rightarrow T$ is a function mapping an entity to its type,  $G_i$ is valid if for each entity v $\in V_i$, $\tau(v) \in$ T, and for each edge (v, v$^\prime$, l) $\in E_i$, ($\tau$ (v), $\tau$ (v$^\prime$), l)$\in$ E. We extend the definition of OPM in paper \cite{ChenENN15}. We will give some definitions here.

{\bf Definition 1} (Open Provenance Model$^+$ (OPM$^+$)) is an extension of OPM recording how is a piece of data derived, which is defined by a triple $<$T, L, G$>$:

\begin{itemize}
	
	\item T is the vertex types: agent (Ag), artifact (A), process (P) and attribute (Att). As shown in Fig \ref{fig2}, the artifact is represented by the shape of oval, which is an object or a piece of data, such as ``$homework_1$", ``\emph{comments}"; A process is an operation performed on a piece of data, such as ``\emph{submit}" and ``\emph{review}"; An agent is a topic that supports operations including ``\emph{$user_1$}" and ``\emph{professor}".   
	
	\item L is the relationship labels: used (u), wasGeneratedBy (wgb), wasControlledBy (wcb), wasTriggeredBy (wtb), wasDerivedFrom(wdf) and hasAttributes (ha).  Each edge in a provenance graph will be marked as one of these labels. Labels describe the relationships between vertices.
	
	\item G is a labelled DAG, where G = $<$V, E$>$, E defines the allowable relationships between the elements, E = \{ (P, A, used), (A, P, wgb), (P, Ag, wcb), (A, A, wdf), (P, P, wtb), (Ag, Att, ha), (P, Att, ha), (A, Att, ha) \}
	
\end{itemize}

{\bf Definition 2} (Provenance Path). A path p = \{ ($v_i$, $v_{i+1}$, ... $v_{i+n}$) $|$ n$\ge$2\}, starting from $v_i$ and ending at $v_{i+n}$, which is a collection of vertices that forms a line in a provenance DAG. In addition, if all the vertices in the path are connected by cause edges , it is a \emph{directional provenance path} indicating that all operations along the \emph{directional provenance path} occur in chronological order. The \emph{general provenance path} may contain effect edges where processes recorded in a \emph{general provenance path} may not occur in chronological order.

\emph{ (XPath Symbols).}   XPath is a query language defined by the World Wide Web Consortium (W3C) for selecting nodes from an XML document. XPath can be used to compute values (e.g., strings, numbers, or Boolean values) from the content of an XML document. Provenance paths are defined over XPath. A directional provenance path can be defined as ($v_i$//$v_j$/$v_k$), describing a provenance path that starts from  $v_i$ and ends by $v_k$, including $v_j$ between $v_i$ and $v_k$.

There are many ways to define the directional provenance path in Xpath can be summarized as follows: “//$v_i$/$v_j$/$v_k$”, which describes a provenance path from $v_i$ to $v_k$ and pass through a node $v_j$ ,and “//$v_i$//$v_j$/$v_k$” ,which queries the node $v_k$ in all $v_j$ nodes beneath $v_i$ , and “//$v_i$//$v_j$[@Attri=’Attri’]/$v_k$”, which queries the node $v_k$ in all $v_j$ nodes with the ‘Attri’ attribute and beneath the node $v_j$.Take the right picture in Figure 3 as an example, we use Xpath rules to traverse the nodes shown in Tab.\ref{tab1}

The keyword is used to distinguish direction and general paths.

\begin{table}
	\centering
	\caption{Xpath Query Usage}
	
	\begin{tabular}{|l|l|l|}
		\hline
		
		Syntax & Usage sample & Description \\ \hline
		//vi/vj/vk &  //o1v2/replace1/o1v1/upload1/au1 & A Query path from o1v2 to  au1 and goes through\\ 
		& & nodes replace, o1v2, upload1 and au1. \\ \hline
		//vi//vj/vk &  //o1v2//o1v1 & Query all o1v1 nodes beneath o1v2 node\\ 
		& & (only one path from o1v2 to o1v1) \\ \hline
		//vi//vj[@Attri=’Attri’] &  //o2v2//review1[@Attri=’Attri’] & Query the review1 node with \\ 
		& &the ‘Attri’ attribute beneath the o1v2 node \\ \hline
		//vi//vj[@Attri=’Attri’]/vk &  //o2v2//review1[@Attri=’Attri’]/au2 & Starting with o2v2, query an au2 node beneath\\ 
		& & review1 with the attribute ‘Attri’ \\ \hline
		
	\end{tabular}
	\label{tab1}
\end{table}

\begin{figure}[ht]
	
	\centering
	$\begin{array}{cc}
	\includegraphics[scale=0.30]{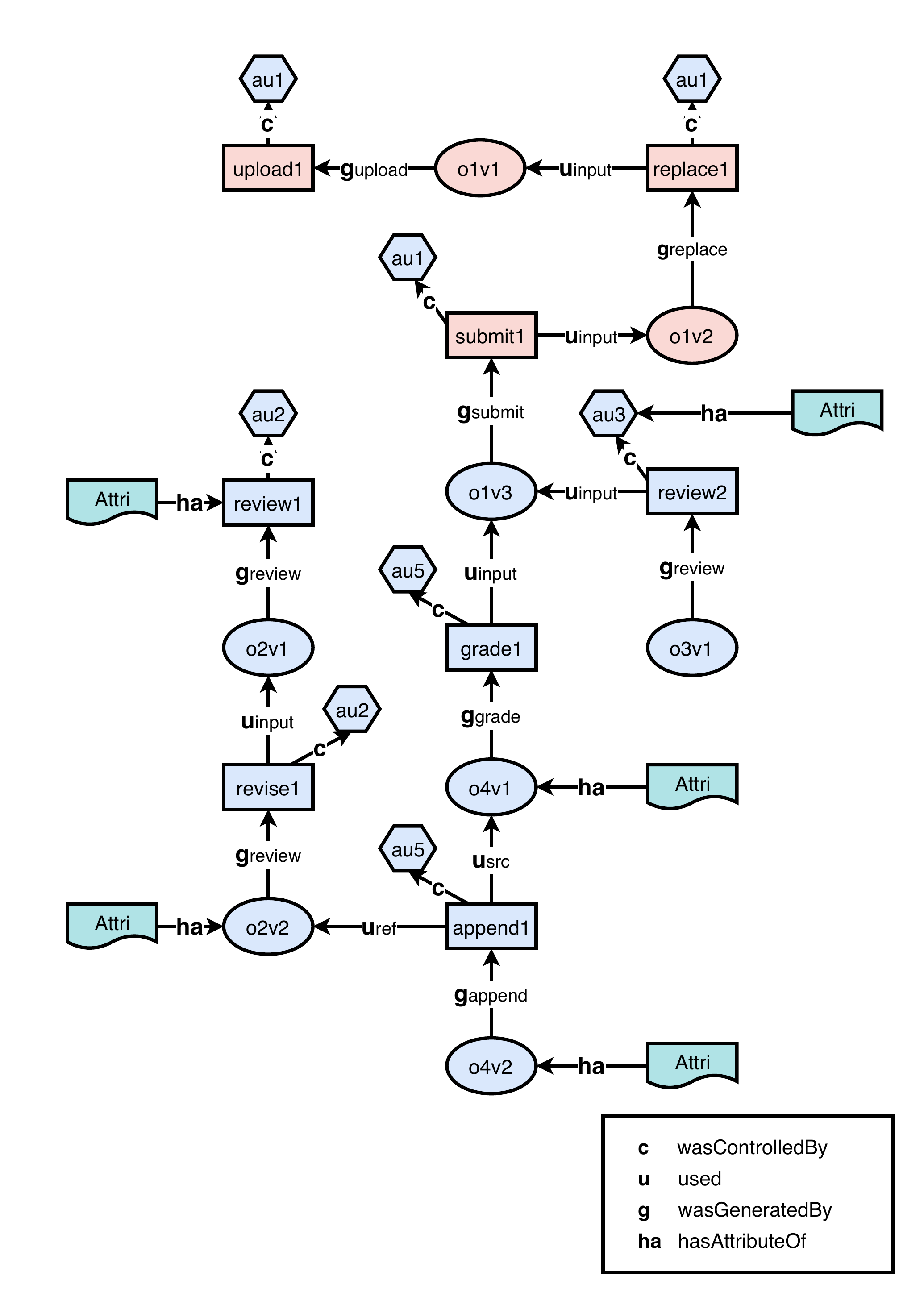} &
	\hspace{-10pt}
	\includegraphics[scale=0.30]{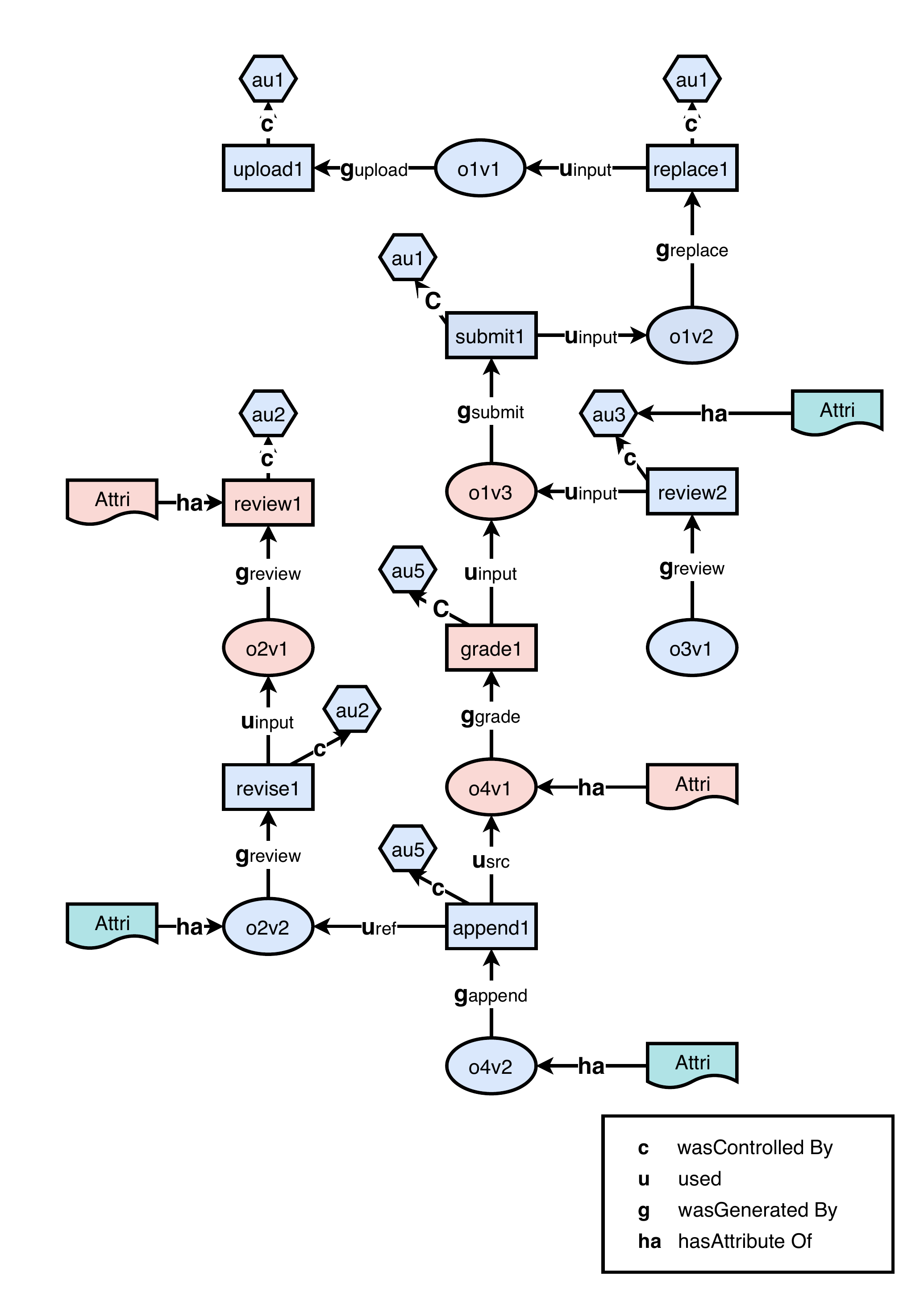}
	
\end{array}$

\caption{Sample Provenance Path A and B}
\label{fig4}
\end{figure}

\begin{itemize}
	\item $(TypedV_{P}^{\prime}(G_i, upload)// TypedV_{P}^{\prime}(G_i, submit))$ 
	
	The first example will match any  provenance path that starts from a \emph{Process} vertex named $upload1$ and ends at a \emph{Process} vertex named $submit1$. Since it is a directed path, all nodes in the path are connected by \emph{CE}. In Fig.\ref{fig4}, the example expression matches the the provenance path described, which can also be represented as follows:
	
	$\rightarrow$ (upload1, o1v1, replace1, o1v2, submit1): $<$upload1, o1v1, $g_{upload}$$>$, $<$o1v1, replace1, $u_{input}$$>$, $<$replace1, o1v2, $g_{replace}$ $>$, $<$o1v2, submit1, $u_{input}$  $>$.
	
	\item $(TypedV_{A}^{\prime}(G_i, o2v1)// TypedV_{A}^{\prime}(G_i, o4v1))$ 
	
	The second example will match any provenance path that starts from a \emph{Artifact} vertex named $o2v1$ and ends at a \emph{Artifact} vertex named $o4v1$. unlike directed path, edges at either end of the path can contain \emph{EE}. In the Fig.\ref{fig4}, the example expression matches the illustrated the provenance path, which can also be represented as follows:
	
	$\rightarrow$ (o2v1, review1(attri), o1v3, grade1, o4v1(attri)): $<$o2v1, review1, $g_{review}$$>$, $<$review1, o1v3, $u_{input}$$>$, $<$o1v3, grade1, $u_{input}$$>$, $<$grade1, o4v1, $g_{grade}$$>$.
	
	\item $(TypedV_{Ag}^{\prime}(G_i, au1), \backslash v+, TypedV_{A}^{\prime}(G_i, o1v2))$

	The last example will match any provenance path that starts from an \emph{Agent} vertex named $au1$ and ends at a \emph{Artifact} vertex named $o1v2$. In the given example graph, there are three coincident paths.
	
	$\rightarrow$ (au1, upload1, o1v1, replace1, o1v2): $<$au1, upload1, c$>$, $<$upload1, o1v1, $g_{upload}$ $>$, $<$o1v1, replace1, $u_{input}$ $>$, $<$replace1, o1v2, $g_{replace}$ $>$; 
	
	$\rightarrow$ (au1, replace1, o1v2): $<$au1, replace1, c$>$, $<$replace1, o1v2, $g_{replace}$ $>$; 
	
	$\rightarrow$ (au1, submit1, o1v2): $<$au1, submit1, c$>$, $<$submit1, o1v2, $u_{input}$ $>$.

\end{itemize}

{\bf Definition 3} (Subgraph). Let G=$<$V, E$>$, S=$<$V$^\prime$, E$^\prime$ $>$. S is a subgraph of G, if V$^\prime$ $\subseteq$ V, E$^\prime$ $\subseteq$ E. A subgraph is a set of vertices \{ ($v_j$, $v_{j+1}$, ... $v_{j+n}$) $|$ n$\ge$2\} in a provenance DAG. A partition \emph{P} of  \emph{G} is a connected subgraph of \emph{G}.\\

\begin{figure}[ht]
	\centering
	$\begin{array}{cc}
	\includegraphics[scale=0.30]{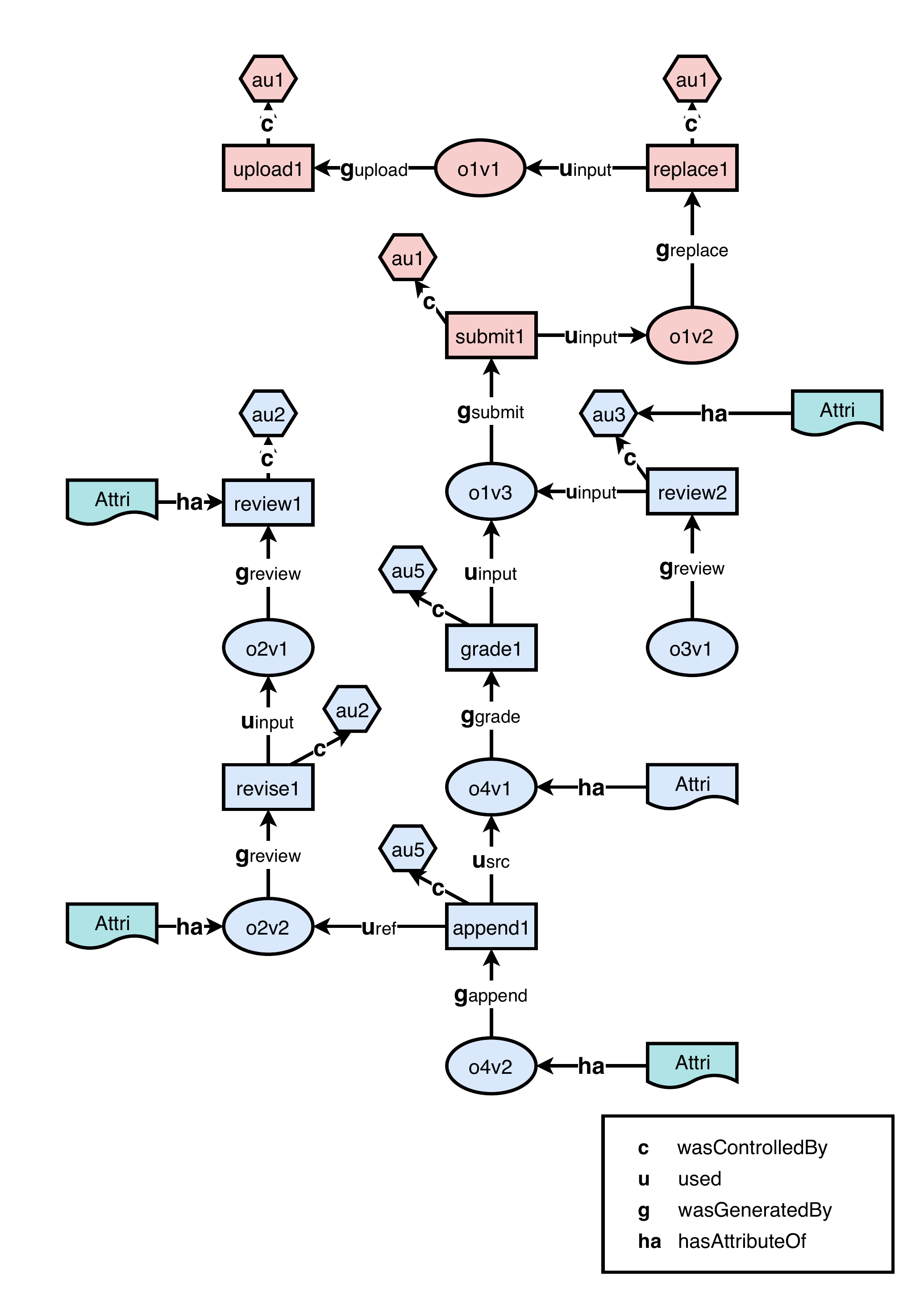} &
	\hspace{-10pt}
	\includegraphics[scale=0.30]{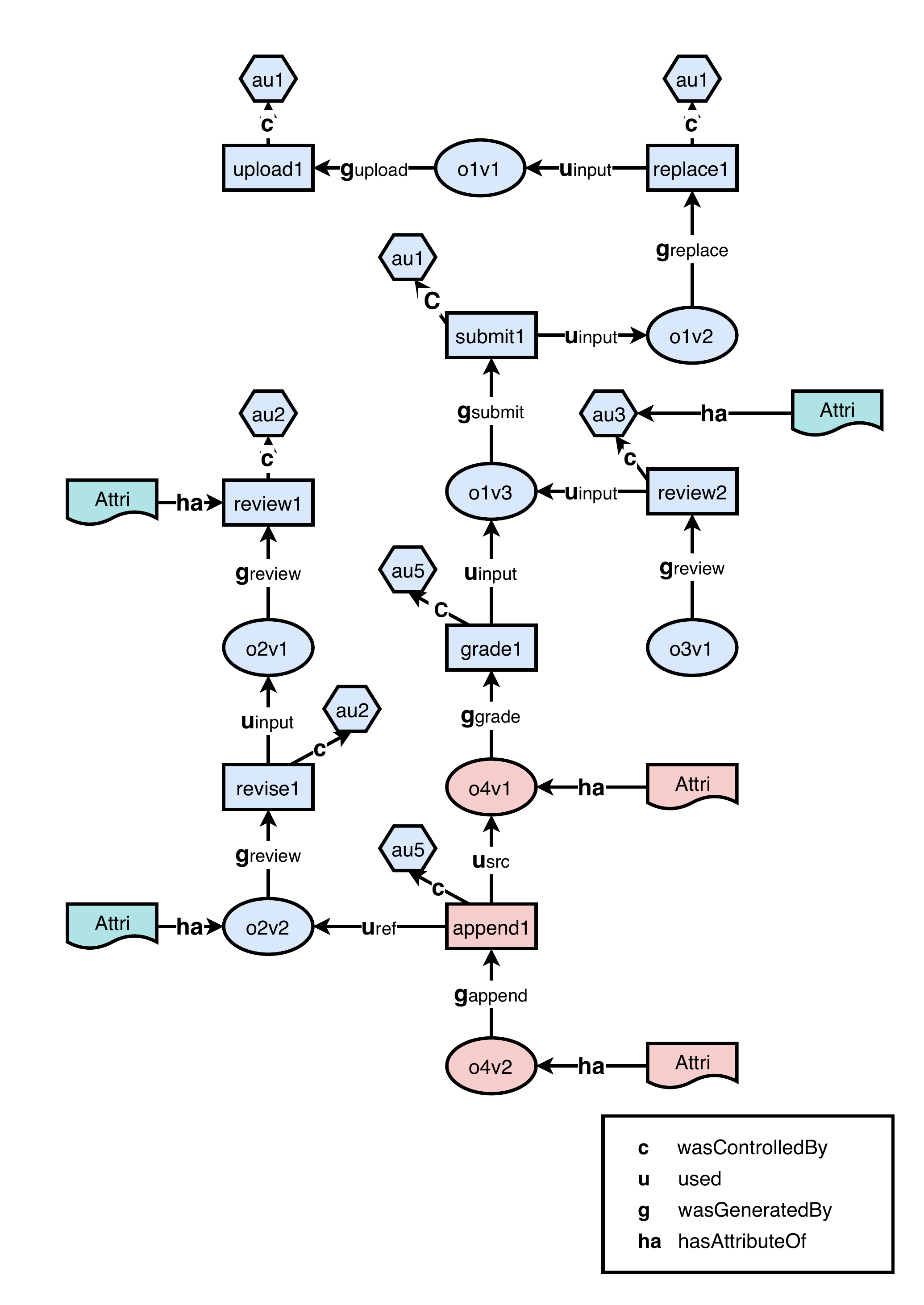}

	\\
\end{array}$
\caption{Sample Subgraph A and B}
\label{fig6}
\end{figure}

We define subgraphs by specifying vertex expressions.  A subgraph consisting of  a given vertex is the target of a policy. The policy defines a subgraph by nominating a starting node and an ending node of the subgraph as the subgraph ($v_i$// $v_j$). A subgraph can start or end at endpoints of a provenance graph regardless of what those vertices are. Hence, in PACLP language, a terminal of a provenance is expressed as $\triangleleft$ which is a starting point of a provenance DAG and $\triangleright$ which is an ending point of a provenance DAG. Specifically, subgraphs ($v_j$/following::*) are graphs that start at the beginning vertex of a provenance graph and end at the Vertex $v_j$. Subgraph ($v_i$/preceeding::*) is that starting at $v_i$ and finishing at the end vertex of a provenance graph.  

Fig.\ref{fig6} highlights a subgraph ($upload//submit1$) starts with  ``upload 1" and ends with ``submit 1". Between the two terminals, it summarizes all the nodes, including vertices of the type of \emph{Agent}. The subgraph ($o4v1$/following::*) is another example between vertex $o4v1$ and the end of the provenance DAG . 

\section{Partition-based Access Control Language on Provenance (PACLP)}
A Partition-based Access Control Policy Language is proposed to extend the access control language \cite{CadenheadKKT11}, enabling a policy to allow or prohibit access to partitions. First, language item PACLP tailored under the  OPM$^+$  stores the attributes that support more fine-grained policies. Second, an \emph{XPath} representation of the provenance partitions is presented to determine which collection of nodes in the source DAG can be accessed.  Therefore, the access to available information could be maximized, rather than hiding the entire graph to protect partially unavailable attributes in a provenance graph. Third, as provenance partitions could provide clues for data sensitivity and vulnerability, the provenance partitions could be employed as conditions in policy.

\begin{figure}[thb]
	\vspace{-0.3cm}
	\centering
	\includegraphics[scale=0.33]{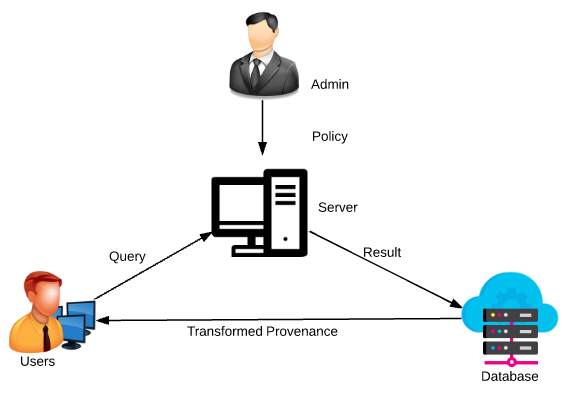}
	\vspace{-0.5cm}
	\caption{System Model} 
	\label{fig11}
\end{figure}

As shown in Fig.\ref{fig11}, the system model consisted of four parts: \emph{Administrator}, \emph{Server}, \emph{Users}, and \emph{Database}. \emph{Users} send queries to the \emph{Server} to access provenance stored in a database. \emph{Administrator} generates access control policies and sends them to the \emph{Server}.  \emph{Server} collects policies from administrators and (optional) data producers, and will generate results based on the policies and delivers the results to the database when receiving queries.
\emph{Database} transforms the target provenance graph based on the results to hide unavailable partitions and sends it to users. Details about the model are discussed bellow.
\begin{figure}[thb]
	\vspace{-0.3cm}
	\centering
	\includegraphics[scale=0.31]{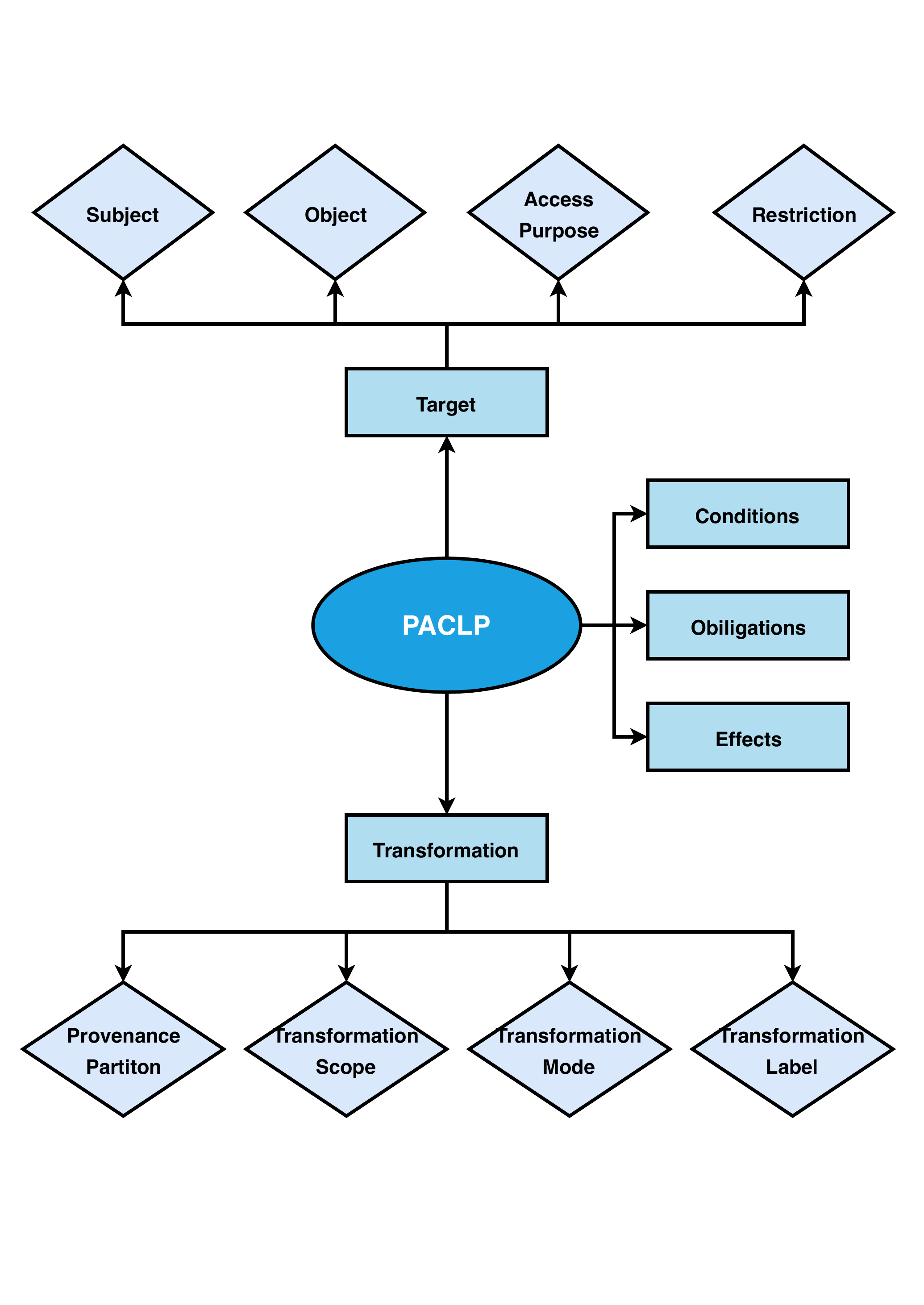}
	\vspace{-0.5cm}
	\caption{PACLP Schema} 
\end{figure}
\label{fig9}

\subsection{Language Items}
Our provenance access control policy PACLP is tailored over XACML syntax consisting of Target, Condition, Obligation, Effect, and Transformation items. Each item contains one or more tags. To support more fine-grained control policies, the PACLP constructed over the OPM$^+$ attaches attributes sets to main entities. In addition, the provenance partitions are employed as conditions to confine applicable policies to determine whether the partitions are accessible. 

Transformation is an important item in policy, indicating how the provenance graphs should be transformed to hide sensitive information. Transformation specifies how each provenance partition should be processed by replacement or deletion. The transformation elements consist of four items, including \emph{provenance partitions}, \emph{transformation scope}, \emph{transformation mode}, and \emph{ transformation label}. First, we demonstrate how to define these provenance partitions in a provenance access control policy and give some examples. 

\begin{itemize}
	\item{\sf Vertices}. This will collect one or more types of vertices in a provenance DAG. These can be Agent, Process, Artifact, or Attributes. Data owners may allow read operations but keep the executor anonymous. 
	As shown below, the agent vertices are collected with values  of ``wasGradedBy" or ``Graded". 
	\end {itemize}
	\makeatletter\def\@captype{table}\makeatother
	%
	
	\noindent \fbox{\parbox{\textwidth}{\ttfamily{
				vertices ($TypedV_{Ag} (G_i)$) \\ 
				vertices ($Typed_P^{\prime} (G_i, wasGradedBy|Graded)$)} }}\\

	\begin{itemize}
		
		\item{\sf Provenance Path}. This is a vertex line from DAG. Two types of source paths are defined based on the directions of the edges : directed path and general path. In a directed path, only edges connect the vertices from the origin to the destination; In general paths, there can exist effect edges between the two endpoints. The following directed path example is from node ``wasGradedBy" or ``Graded" to node ``wasSubmittedBy" or ``Submit".  Particularly, in a directed path the process from the original to the target is listed in chronological order, as all nodes are linked by cause edges.

	\end{itemize}
	%
	
	\noindent \fbox{\parbox{\textwidth}{\ttfamily{
				directed ($TypedV_{Ag}^{\prime} (G_i, wasGradedBy|Graded)//$b+\\
				$TypedV_{Ag}^{\prime} (G_i, wasSubmittedBy|Submit)$)} }}\\
	
	\begin{itemize}
		
		\item{\sf Subgraph}. A subgraph is a collection of vertices with a specified origin and/or destination, and can be represented as a subgraph ($v_i$ $\backslash v+ v_j$). The  first example of the subgraph below defines all operations performed in the provenance graph $G_i$ in 2016. Another example represents a partition from the vertex with a given value to the end of the entire graph, representing all the operations that have occurred in the graph since 2016.\\ 
	\end{itemize}
	%
	
	\noindent \fbox{\parbox{\textwidth}{\ttfamily{
				subgraphs $(AttV_{P} (G_i, 1/1/2016)//AttV_{P} (G_i, 31/12/2016))$ \\
				subgraphs $(AttV_{P} (G_i, 1/1/2016)/following::*))$}   }}\\ 
	\\
	
	\emph{Transformation scope} defines the scope of the node used for transformation, which accepts three possible values: Original, Conjunction, and Extension. Original means that the vertices defined by access control policies do not extend to other vertices in the given provenance graph. Conjunction
	indicates that a set of vertices should be integrated with the connection nodes in the same category as VCD. To facilitate graph transformation, the VCD lists categories of nodes and corresponding labels. When the graph is transformed, the label replaces the  removed vertex. If the cluster's neighbor nodes belong to the same category, the cluster should expand to include those neighbor nodes. Extension is a function that returns a set of clusters in a given provenance graph. For a given cluster of vertices, all vertices in the provenance graph belong to the same category and should be collected as a set of clusters, regardless of  whether they are connected to a given cluster.
	

	\emph{Transformation mode} indicates how to handle vertices that can be accessed or collected by provenance access control policies. Two possible modes are ``replace" and ``remove". For Replace mode, when transforming a new provenance graph, replace the vertex cluster with the label pointing to the VCD. Labels are summary terms for vertex categories. For Remove mode, it removes clusters of vertices specified by access control policy, and edges appear outside.
	
	\emph{Transformation labels}, for the  Original dependency,  when a cluster of nodes is removed or replaced, the original dependency means that the two edges beside a removed node are merged by referencing to the  Edge Merging Table to maintain the original dependency. For the Fault dependency, it means removing the original label and replacing it with "wasCausedBy" to prevent the label showing clues to remove vertices.

	The sample \emph{Transformation} item defines two provenance partitions to be transformed. The first one is a \emph{subgraph} that starts at an \emph{Artifact} vertex \emph{o3v1} and ends at \emph{Artifact} vertex \emph{o8v1}. As the transformation scope is ``original", the subgraph does not contain any other vertices. Since the transform mode is "replace". It should be replaced with a label. In addition, all the connection edges of the node cluster should be changed to "wasCausedBy". The second hidden partition is a \emph{Process} vertex Submit $|$ was SubmittedBy. If adjacent nodes belong to the same category as defined by the VCD, they are included in a partition, which is deleted according to the transformation mode. For edges, keep the original edges, or merge them by referring to the Edge Merging Table.
	
	\noindent \fbox{\parbox{\textwidth}{\ttfamily{
				$<$Transformation$>$\\
				$<$partition$>$\\
				subgraphs ($TypedV_{A}^{\prime} (G_i, o3v1)//TypedV_{A}^{\prime} (G_i, o8v1)>$) \\
				$<$/partition$>$\\
				$<$scope$>$ original $<$/scope$>$\\
				$<$mode$>$ replace $<$/mode$>$\\
				$<$label$>$ false dependency $<$/label$>$\\
				$<$partition$>$ \\
				vertices ($TypedV_{P}^{\prime} (G_i, Submit|wasSubmittedBy)$)\\
				$<$/partition$>$\\
				$<$scope$>$ conjunction $<$/scope$>$\\
				$<$mode$>$ remove $<$/mode$>$\\
				$<$label$>$ original dependency $<$/label$>$\\
				$<$/Transformation$>$}
	}}\\
	~\\

	\subsection{Evaluation}
	
	To evaluate the performance of the PACLP, the experiment is designed to simulate the policy generation, performed onto a virtual machine with 16GB memory and 3.40 GHz CPU. The 20 sample provenance graphs and 200 policy conditions and tags are generated. We select 300 random provenance partitions from the sample provenance graphs. 
	We count the numbers of provenance partitions under PACLP and LPAC. Obviously, in the storage entity attributes of provenance model, compared with other other policy languages, PACLP can express more random sample provenance partitions. From the figures, we can see PACLP is good at to describe more complicated provenance partitions with more nodes. Then, we implement the process of merging the results of various policies. The three scenarios selected are (1) all 20 sample policies are policies with the effect of \emph{Abosulte Permit} 
	(2) all 20 sample policies with the effect of \emph{Deny}, (3) 20 sample policies with all effects are randomly selected. We count the time span to simulate the process.

	\begin{figure*}[ht]
		\vspace{-10pt}
		\centering
		$\begin{array}{cc}
		\includegraphics[width=5.2cm]{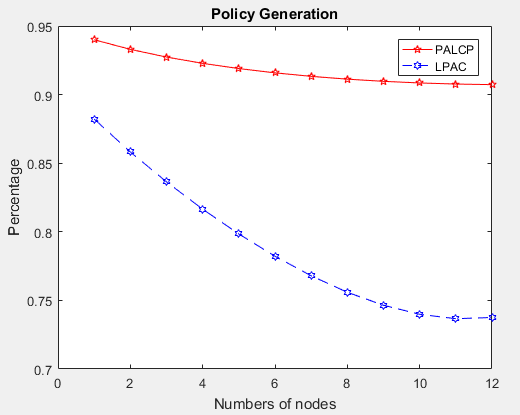} &
		\hspace{15pt}
		\includegraphics[width=5.32cm]{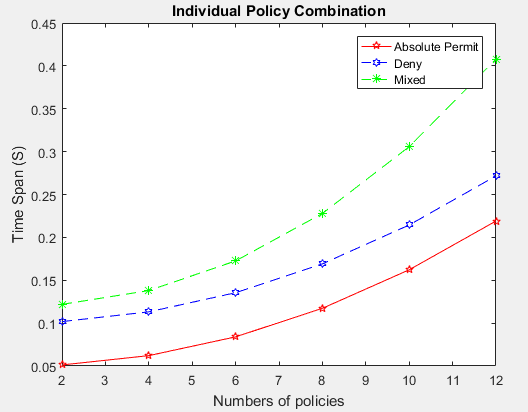} 
		\\
		\end{array}$
		\caption{left: Comparison of express ability of PACLP and LPAC. right: The time span of policy results combination with different number of policies.}
		\label{fig:ourlmodels}
	\end{figure*}

	\section{Conclusion}

	In this paper, we propose a fine-grained provenance access control language PACLP under extended OPM storing attribute sets to extend the exiting languages. Various types of provenance  partitions are defined over XPath to determine which nodes in the source DAG can be accessed. Our provenance access control language aims to define which partial graph can be accessed or denied under conditions and restrictions. This fine-grained access control policy model not only hides all sensitive vertices and edges in the provenance graph, but also maximizes accessible qualifying information.

\end{document}